\documentclass[smallextended]{svjour3}       
\smartqed  
\usepackage{times}
\usepackage{mathtools}
\usepackage{amsmath}
\usepackage{amssymb}
\usepackage[normalem]{ulem}
\usepackage[unicode=true,pdfusetitle,
 bookmarks=true,bookmarksnumbered=false,bookmarksopen=false,
 breaklinks=true,pdfborder={0 0 1},backref=false,colorlinks=true]
 {hyperref}
\hypersetup{
 pdftex,pdfborderstyle=,allcolors=blue}
 
\usepackage{color}
\usepackage{url}
\usepackage{graphicx}
\usepackage{amsmath}
\usepackage{amssymb}
\usepackage{mathtools}

\def\rd{{\rm d}}
\def\vb{{\bf b}}

\def\vx{{\bf x}}
\def\mA{{\bf A}}
\def\mD{{\bf D}}
\def\vF{{\bf F}}
\def\mJ{{\bf J}}
\def\vX{{\bf X}}

\def\vy{{\bf y}}
\def\vf{\mbox{\boldmath$f$}}
\def\vv{{\bf v}}
\def\vu{{\bf u}}

\def\dbar{{\mathchar'26\mkern-12mu \rd}}

\begin{document}

\title{{Bivectorial Nonequilibrium Thermodynamics:}}

\subtitle{{Cycle Affinity, Vorticity Potential, and Onsager's Principle
}}


\author{Ying-Jen Yang         \and
        Hong Qian 
}


\institute{Ying-Jen Yang \at
              Department of Applied Mathematics, University of Washington,
Seattle, WA 98195-3925, USA \\
              \email{yangyj@uw.edu}           
           \and
           Hong Qian \at
            Department of Applied Mathematics, University of Washington,
Seattle, WA 98195-3925, USA \\
              \email{hqian@uw.edu}
}

\date{Received: date / Accepted: date}

\maketitle

\begin{abstract}
{We generalize an idea in the works of Landauer and Bennett on computations, and Hill's in chemical kinetics, to emphasize the importance of kinetic cycles in mesoscopic nonequilibrium thermodynamics (NET).} For continuous
stochastic systems, a NET in phase space is formulated in terms of
cycle affinity $\nabla\wedge\big(\mD^{-1}\vb\big)$ and vorticity {potential 
$\mA(\vx)$ of} the stationary flux $\mJ^{*}=\nabla\times\mA$.
Each bivectorial cycle couples two transport processes represented
by vectors and gives rise to Onsager's notion of reciprocality; 
the scalar product
of the two bivectors $\mA\cdot\nabla\wedge\big(\mD^{-1}\vb\big)$
is the rate of local entropy production in the nonequilibrium steady
state. An Onsager operator that relates vorticity to cycle affinity is
introduced. 
\end{abstract}

\section{Introduction.}

Nonequilibrium thermodynamics (NET) pioneered by L. Onsager \cite{onsager_reciprocal_1931}
is concerned with a diverse array of macroscopic physical and chemical
processes: mass transport, heat conduction, chemical reactions, etc.
A unified treatment in continuous systems was developed in the 1960s
\cite{groot_non-equilibrium_2011}. In recent years, introducing a
NET of mesoscopic stochastic dynamics in its phase space has provided
a more fundamental formulation in which the different physical and
chemical fluxes are all represented by a single probability flux.
Positivity of mean entropy production can be mathematically demonstrated,
and large deviation fluctuation theorems were discovered \cite{lebowitz_gallavotti-cohen-type_1999}.
The local equilibrium assumption required in \cite{groot_non-equilibrium_2011}
does not enter the stochastic theory \emph{per se} until its application
when constitutive models for real world processes are required. In
discrete-state systems, {\em cycle flux} and {\em cycle affinity}
play fundamental roles in its NET; the initial idea goes back to \cite{hill_free_2012}.
See \cite{qian_entropy_2016} for a recent synthesis.

{
\section{Cycle completion and Irreversibility.}

Consider the following thought experiment. Suppose we have a discrete-state Markov process and would like to determine whether its stationary process is detailed balanced or not, \emph{i.e.} in equilibrium or not.  We run the process with a trajectory $i_{0} i_{1} \cdots i_{k}$ and measure the dissipation

\begin{equation}
\mathcal{Q}(i_{0} i_{1} \cdots i_{k})=\ln\left(\frac{q_{i_{0}i_{1}}q_{i_{1}i_{2}}\cdots q_{i_{k-1}i_{k}}}{q_{i_{1}i_{0}}q_{i_{2}i_{1}}\cdots q_{i_{k}i_{k-1}}}\right)
\end{equation}where $q_{ij}$ is the (unknown) conditional transition rate from state $i$
to $j$.
Komogorov's cycle condition states that the process admits a detailed balanced steady state if and only if $\mathcal{Q}(\sigma)=0$ for all cycles $\sigma$, \emph{e.g.} $i_{0} i_{1} \cdots i_{k}i_0$. 
Therefore, we shall examine $\mathcal{Q}$ everytime the process completes a cycle, and if $\mathcal{Q}=0$, we continue to run the process for another cycle. We can draw the conclusion either when we encouter a cycle with nonzero $\mathcal{Q}$, signifying detailed balance  broken, or we complete all cycles and find that all cycles are reversible with zero dissipation.

In the thought experiment above, we see that it is essential to finish cycles to determine whether a system admits detailed balance or not. We can't gain useful information before the process completes a full cycle. 
This is because for a trajectory one-step before the completion of a cycle, say $i_{0},i_{1},\cdots i_{k}$,
with all distinct states, there's always a possibility that the last step balances out the probability difference, \emph{i.e.} $\mathcal{Q}(i_{0} i_{1} \cdots i_{k})+\mathcal{Q}(i_{k} i_0)=0$, giving us  a reversible cycle with no dissipation.} We shall
call this observation Landauer-Bennett-Hill (LBH) principle: In the
theory of computation, R. Landauer applied the second law of thermodynamics 
to point out the necessary accompanied heat dissipation of ``erasing
one bit'' \cite{landauer_irreversibility_1961}; C. H. Bennett then used
Landauer's principle to argue that it is the last step of ``erasing
bits'' in a cyclic Maxwell demon that ``saves'' the second law
\cite{bennett_notes_2003}.  Independently in the theory of cycle kinetics
driven by chemostatic chemical potential, T. L. Hill introduced the
concept of {\em cycle completion} \cite{hill_stochastics_1975}
and argued that cycles in mesoscopic NET are more fundamental than
transitions \cite{hill_free_2012}. The notion of ``erasing one bit''
of Landauer's and Bennett's matches exactly the idea of ``completing
one cycle''.

Parallel to the cycle representation of discrete-state Markov processes
which has been extensively studied \cite{kalpazidou_cycle_2006},
here we present a cycle representation for the NET of continuous Markovian
stochastic dynamics in its phase space{ the whole Euclidean $\mathbb{R}^{n}$,  and discuss
how the LBH principle comes in. The $n=2$ (and $n=3$ implied) case,
in which a vector potential arises, has been investigated by one of the us \cite{qian_vector_1998}. However, the generalization to $n>3$ systems is nontrivial. One of the difficulties is that,  for systems with $n>3$, vector calculus is no long sufficient since it is not possible to represent vorticity by a vector through the right-hand rule: there are more than one dimensions in the ``thumb'' direction.} It turns out that both the cycle flux
and cycle affinity in the continuous system are bivectors (see Appendix),
which can be represented by their skew-symmetric $n\times n$ matrices components.  
 
More importantly, while nonequilibrium steady state (NESS) cycle flux as
a kinematic concept is nonlocal and requires highly nontrivial computation,
the cycle affinity that quantifies NET thermodynamic driving force is locally
determined and completely independent of the kinematics. The bivectorial
nature of a cycle reflects the coupling between any two transport processes, visualizable as two vectors in $\mathbb{R}^n$,
$X_{i}$ and $X_{j}$. This further implies the fundamental importance
of cycles: a nonequilibrium device converts the force in $X_i$
to the transport in $X_j$, the two dimensions form a cycle and
{reciprocality} naturally follows. {{In an equilibrium
steady state, any clockwise and counter-clockwise fluxes on a cycle 
are exactly the same; thus Onsager's reciprocal symmetry for NET in the linear regime follows.}}
The bivector formalism helps establishing a clear physical picture
and the mathematical representation of reciprocality in NET envisioned
by Onsager.  

\section{Continuous Markov Processes.}

Consider a mesoscopic system represented by a continuous Markov process
with diffusion matrix $\mD(\vx)$ and drift $\vb(\vx)$, $\vx\in\mathbb{R}^{n}$. {Here we consider the dynamics on the whole Euclidean $\mathbb{R}^n$.}
The stochastic dynamics is described by a time-dependent probability
density function $p(\vx,t)$ that follows the Fokker-Planck equation
(FPE) 
\begin{equation}
\partial_{t}p(\vx,t)=-\nabla\cdot\left[\vb(\vx)p(\vx,t)-\mD(\vx)\nabla p(\vx,t)\right].\label{FPE}
\end{equation}
With Ito's calculus, this has a corresponding trajectory-based stochastic
differential equation, 
\begin{equation}
\rd\mathbf{X}_{t}=\left[\vb\left(\mathbf{X}_{t}\right)+\nabla\cdot\mD\left(\mathbf{X}_{t}\right)\right]\rd t+\mathbf{\Gamma}\left(\mathbf{X}_{t}\right)\rd\mathbf{W}_{t}\label{eq: SDE}
\end{equation}
where $\mD=\mathbf{\Gamma}\mathbf{\Gamma}^{\mathsf{T}}/2$,
$\left(\nabla\cdot\mD\right)_{i}=\sum_{j=1}^{n}\partial_{j}D_{ji}$,
and $\mathbf{W}_{t}$ is the $n$ dimensional Brownian motion. We've
denoted $\partial_{j}$ as the partial derivative with respect to
$x_{j}$.

With Eq. \eqref{FPE}, the probability flux at $t$ is given by 
\begin{equation}
\mJ(\vx,t)=\vb(\vx)p(\vx,t)-\mD(\vx)\nabla p(\vx,t),\label{eq: Probability flux}
\end{equation}
and the notion of ``probability velocity'' can be introduced as
$\mathbf{v}(\vx,t)=\mJ\left(\vx,t\right)/p(\vx,t).$ In the stationary
state, we have an invariant probability density $\pi(\vx)$, a divergence-free
stationary flux 
\begin{equation}
\mJ^{*}(\vx)=\vb(\vx)\pi(\vx)-\mD(\vx)\nabla\pi(\vx),
\end{equation}
and a ``stationary probability velocity'' $\mathbf{v}^{*}(\vx)=\mJ^{*}\left(\vx\right)/\pi(\vx).$
An equilibrium corresponds to detailed balanced condition: $\mJ^{*}(\vx)=0=\mathbf{v}^{*}(\vx)$.

\section{Infinitesimal change and cyclic change of thermodynamic quantities.}

Mesoscopic thermodynamics concerns the rate of change, production
and dissipation of mainly three thermodynamic quantities: the (stochastic)
Shannon entropy $S(\vx,t)\coloneqq-\ln p(\vx,t)$, the \emph{nonequilibrium
potential energy $\Phi(\vx)\coloneqq-\ln\pi(\vx)$, }and the free
energy $F(\vx,t)\coloneqq\Phi(\vx)-S(\vx,t)$ \cite{seifert_stochastic_2012,yang_unified_2020}.
Their infinitesimal (stochastic) change along $\vX_{t}$ from $t$
to $t+\rd t$ can be expressed as \begin{subequations}\label{production of Phi, S, and F}
\begin{align}
\mathrm{d}\Phi(\vX_{t}) & =\nabla\Phi(\vX_{t})\circ\mathrm{d}\mathbf{X}_{t}\label{eq: dPhi}\\
\mathrm{d}S(\vX_{t},t) & =\partial_{t}S(\vX_{t},t)\rd t+\nabla S\left(\vX_{t},t\right)\circ\mathrm{d}\mathbf{X}_{t}\label{eq: dS}\\
\mathrm{d}F(\vX_{t},t) & =-\partial_{t}S(\vX_{t},t)\rd t+\nabla F\left(\vX_{t},t\right)\circ\mathrm{d}\mathbf{X}_{t}.\label{eq: dF}
\end{align}
\end{subequations}Here $\circ$ denotes the Stratonovich midpoint
integration: $\mathbf{u}(\mathbf{X}_{t},t)\circ\mathrm{d}\mathbf{X}_{t}$ is equal to $\mathbf{u}\left(\mathbf{X}_{t}+\tfrac{1}{2}\mathrm{d}\mathbf{X}_{t},t\right)\cdot\mathrm{d}\mathbf{X}_{t}$
which takes care of the extra term in Ito's calculus due to the $\sqrt{\rd t}$
scaling of $\mathrm{d}\mathbf{W}_{t}$ \cite{lebowitz_gallavotti-cohen-type_1999,qian_mesoscopic_2001}. 

The instantaneous production of entropy $\rd S$ has a decomposition
$\rd S=\dbar\mathcal{S}_{\mathrm{tot}}-\dbar\mathcal{Q}$ in terms
of the following two quantities, \begin{subequations}

\begin{align}
\dbar\mathcal{Q} & =\mD^{-1}\vb\circ\rd\mathbf{X}_{t}\label{eq: heat}\\
\dbar\mathcal{S}_{\mathrm{tot}} & =\partial_{t}S\rd t+\mD^{-1}\mathbf{v}\circ\mathrm{d}\mathbf{X}_{t}.\label{eq: Stot}
\end{align}
\end{subequations}
They are the total amount of heat dissipated from the system to the
environment \cite{lebowitz_gallavotti-cohen-type_1999} and the total
entropy production of the system and the environment. Note the important
distinction: Infinitesimal change of a function $A(\vX_{t},t)$ is
$\rd A\equiv A(\vX_{t+\rd t},t+\rd t)-A(\vX_{t},t)$; but there is
no such a function for $\ \dbar\mathcal{B}$ in general. The latter
represents work against a non-conservative force, or a ``source''
term. $\mathcal{B}$, or any calligraphic letter in the present work,
is a path observable.

When the system reaches its NESS, the total entropy production at
the steady state is the difference between the total heat dissipation
$\thinspace\dbar\mathcal{Q}$ and the excess heat dissipation associated
with the change in the nonequilibrium potential $\dbar\mathcal{Q}_{\mathrm{ex}}=-\nabla\Phi\circ\mathrm{d}\mathbf{X}_{t}.$
It is the amount of energy needed to sustain the steady state, called
\emph{housekeeping heat}, 
\begin{align}
\dbar\mathcal{Q}_{\mathrm{hk}} & \equiv\dbar\mathcal{Q}-\dbar\mathcal{Q}_{\mathrm{ex}}=\mD^{-1}\mathbf{v}^{*}\left(\mathbf{X}_{t}\right)\circ\mathrm{d}\mathbf{X}_{t}.\label{eq: dQhk}
\end{align}
From Eqs (\ref{eq: dF}), (\ref{eq: heat}) and (\ref{eq: dQhk}),
one gets the entropy production decomposition $\thinspace\dbar\mathcal{S}_{\mathrm{tot}}=\dbar\mathcal{Q}_{\mathrm{hk}}-\rd F$.
See \cite{yang_unified_2020} for a recent synthesis.

To compute the mean rate of the thermodynamic quantities above, we
consider the infinitesimal change of a ``work''-like quantity $\mathcal{W}$
associated with a ``force'' field $\vf(\vX_{t},t)$, 
\begin{equation}
\thinspace\dbar\mathcal{W}=\vf(\vX_{t},t)\circ\rd\vX_{t}.\label{eq: change in a work-like quantity}
\end{equation}
For a smooth cyclic path $\Gamma:\mathbf{x}(t),0\le t\le\tau$ where
$\mathbf{x}(0)=\mathbf{x}(\tau)=\boldsymbol{\xi}$ in $\mathbb{R}^{n}$,
the cyclic ``work'' can be rewritten {
 by considering a surface $\Sigma$ whose boundary is given by the path $\Gamma$ and using Stoke's theorem \cite{feng_potential_2011}.
\begin{subequations}
\begin{align}
\mathcal{W}(\Gamma) 
& =\oint_{\Gamma}\vf\cdot\mathrm{d}\mathbf{x} = \oint_{\Gamma} \sum_{i=1}^{n} f_i \mathrm{d}x_i\\
& =\int_{\Sigma} \sum_{1\le i<j \le n} \left(\partial_{i}f_{j}-\partial_{j}f_{i}\right)\mathrm{d}x_{i}\wedge\mathrm{d}x_{j} 
=\int_{\Sigma} \nabla\wedge\vf\cdot\mathrm{d}\boldsymbol{\sigma}.\label{eq: Stoke's theorem stopath}
\end{align}
\end{subequations}
In Eq. \eqref{eq: Stoke's theorem stopath}, neither $\nabla\wedge\vf$ nor $\mathrm{d} \boldsymbol{\sigma}$ are vectors in $\mathbb{R}^{n}$; rather they are bivectors, planary objects that have skew-symmetric matrix components with respect to the orthonormal basis $\{\mathbf{e}_i\wedge \mathbf{e}_j,1\le i<j\le n\}$ where $\mathbf{e}_i$ is the unit vector in the $i$th direction of Cartesian coordinate and  $\mathbf{e}_i\wedge \mathbf{e}_j$ is the signed area of the parallelogram spanned by $\mathbf{e}_i$ and $\mathbf{e}_j$.
The $\nabla\wedge\vf=\sum_{i<j} (\partial_{i}f_{j}-\partial_{j}f_{i})\mathbf{e}_i\wedge \mathbf{e}_j$ denotes the ``curl'' of $\vf$, representing the vorticity of $\vf$ by a bivector with skew-symmetric matrix components.
Here, we understand $\mathrm{d}x_{i}\wedge\mathrm{d}x_{j}$ as the $(i,j)$th components of the infinitesimal plane $\mathrm{d}\boldsymbol{\sigma}$ as a bivector.
The dot product in Eq. \eqref{eq: Stoke's theorem stopath} between two bivectors are defined as the half of the Frobenius product between their matrix components. See Appendix for a more detailed introduction in the language of differential form.}

The cyclic changes of the thermodynamic quantities are then given
by, \begin{subequations}\label{cycle production of ther quantities}
\begin{align}
\Delta\Phi(\Gamma) & =-\mathcal{Q}_{\mathrm{ex}}(\Gamma)=0\label{eq: Qex (cycle)}\\
\Delta S(\Gamma) & =-\Delta F(\Gamma)=S(\boldsymbol{\xi},\tau)-S(\boldsymbol{\xi},0)\label{eq: free energy dissipation}\\
\mathcal{Q}(\Gamma) & =\mathcal{Q}_{\mathrm{hk}}(\Gamma)=\int_{\Sigma}\nabla\wedge\left(\mD^{-1}\vb\right)\cdot\mathrm{d}\boldsymbol{\sigma}\label{eq: Q cycle}\\
\mathcal{S}_{\mathrm{tot}}(\Gamma) & =\Delta S(\Gamma)+\mathcal{Q}(\Gamma).\label{eq: Stot cycle}
\end{align}
\end{subequations}
If the path probability of a path $\Gamma$ in a Markov process starts
with the invariant probability as the initial distribution, $\Delta S\left(\Gamma\right)=0$
for all cycles $\Gamma$, over which the total entropy production
equals to the heat dissipation: 
\begin{equation}
\mathcal{S}_{\mathrm{tot}}^{*}(\Gamma)=\mathcal{Q}(\Gamma)=\int_{\Sigma}\nabla\wedge\left(\mD^{-1}\vb\right)\cdot\mathrm{d}\boldsymbol{\sigma}.\label{eq: stochastic Stot in NESS}
\end{equation}
$S$ and $\Phi$ are state functions, but $\mathcal{Q}$ and $\mathcal{S}_{\mathrm{tot}}^{*}$
are not. These are direct consequences of \eqref{production of Phi, S, and F}
and \eqref{eq: heat} in the cycle representation. In fact, Eq. \eqref{eq: stochastic Stot in NESS}
shows that $\nabla\wedge\left(\mD^{-1}\vb\right)\cdot\mathrm{d}\boldsymbol{\sigma}$
is the cyclic entropy production of an infinitesimal cycle $\mathrm{d}\boldsymbol{\sigma}$,
and the total cyclic entropy production is the integral of all the
infinitesimal cycle that tiles the cycle $\Gamma$. {This implies that the bivector $\nabla\wedge\left(\mD^{-1}\vb\right)$
can be interpreted as the \emph{cycle} \emph{affinity} in diffusion, and the Komogorov's cycle condition in diffusion for detailed balanced systems becomes a zero cycle affinity condition $\nabla\wedge\left(\mD^{-1}\vb\right)=\mathbf{0}$. With Poicar\'{e} lemma applied on the contractible $\mathbb{R}^n$, this implies that $\mD^{-1}\vb$ is curl-free if and only if it is a gradient field, with a scalar potential given by $\Phi$.

\begin{figure}
\[
\includegraphics[scale=0.25]{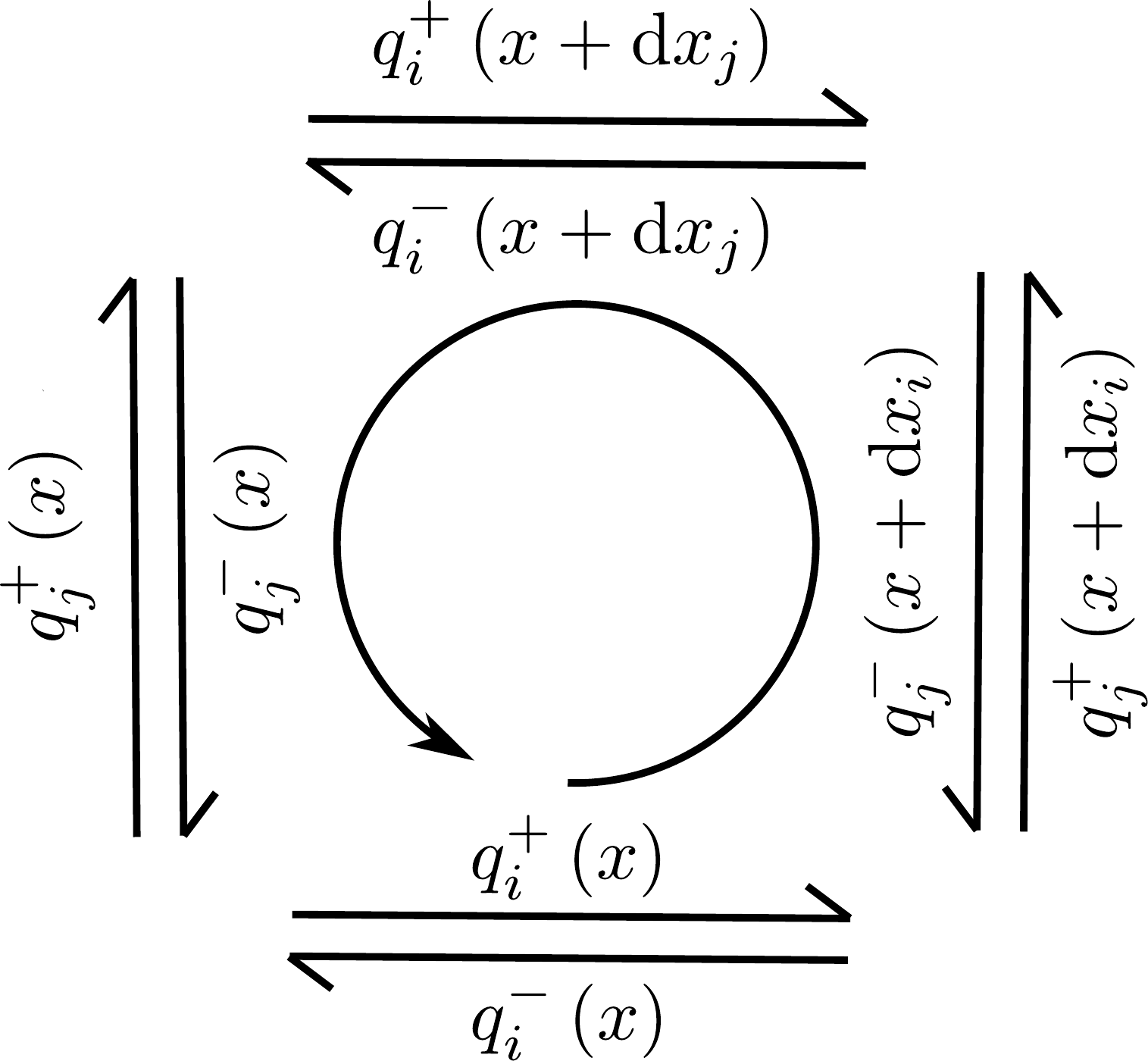}
\]
\caption{Transitions on a infinitesimal square $\rd x_i \wedge \rd x_j$ in a lattice. The eight transition rates $q^{\pm}_{i\ \mathrm{or}\ j}$ determine the cycle affinity. This explains the physical picture of cycle affinity as shown in Eq. \eqref{eq: Q in lattice}.}
\label{cycle affinity} 
\end{figure}

The physical picture of cycle affinity is clear in the discrete lattice picture as shown in Fig. \ref{cycle affinity}.
Focusing on the infinitesimal square of $\rd x_i \wedge \rd x_j$ at $x$, the cycle affinity of the counterclockwise cycle is given by 
\begin{subequations} \label{eq: Q in lattice}
\begin{align}
\mathcal{Q}_{ij} &= \ln\frac{q_{i}^{+}(x)q_{j}^{+}(x+\rd x_{i})q_{i}^{-}(x+\rd x_{j})q_{j}^{-}(x)}{q_{i}^{-}(x)q_{j}^{-}(x+\rd x_{i})q_{i}^{+}(x+\rd x_{j})q_{j}^{+}(x)}\\
 & \sim \left[\partial_{i}\left(\frac{1}{\rd x_{j}}\ln\frac{q_{j}^{+}(x)}{q_{j}^{-}\left(x\right)}\right)-\partial_{j}\left(\frac{1}{\rd x_{i}}\ln\frac{q_{i}^{+}(x)}{q_{i}^{-}\left(x\right)}\right)\right]\rd x_{i}\rd x_{j}.
\end{align}
\end{subequations}
where $\frac{1}{\rd x_j}\ln\frac{q_{j}^{+}}{q_{j}^{-}}$ becomes $\left(\mD^{-1} \vb\right)_j$ from the theory of diffusion. 
We see that the particular order of $\partial_i \left(\mD^{-1} \vb\right)_j - \partial_j \left(\mD^{-1} \vb\right)_i$ is because the increment of $q_j^{+}/q_j^{-}$ in the $i$-th direction flavors the counterclockwise rotation whereas the increment of $q_i^{+}/q_i^{-}$ in the $j$-th direction hinders it.
Their net contribution gives the force on the cycle, the cycle affinity.
}

The mean rate of $\mathcal{W}(t)$ in \eqref{eq: Stoke's theorem stopath}
can be computed following Ito's calculus \cite{qian_mesoscopic_2001},{
\begin{equation}
\dot{\mathcal{W}}\equiv\frac{1}{\mathrm{\rd t}}\mathbb{E}\left[\thinspace\dbar\mathcal{W}\right]=\int_{\mathbb{R}^{n}}\mJ(\mathbf{x},t)\cdot\vf(\mathbf{x},t)\rd V\label{eq: average production rate}
\end{equation}
where $\rd V=\prod_{i=1}^{n}\rd x_{i}$ is the volume of a $n$-dimensional  infinitesimal  cube} and $\mathbb{E}\left[\cdot\right]$
denotes expectation. The mean rates of $\Phi$, $S$, $F$, $\mathcal{S}_{\mathrm{tot}}$,
and $\mathcal{Q}_{\mathrm{hk}}$ can then be obtained by plugging
in the corresponding forces $\vf=\nabla\Phi,\nabla S,\nabla F,\mathbf{D}^{-1}\mathbf{v}$,
and $\mathbf{D}^{-1}\mathbf{v}^{*}$ \cite{yang_unified_2020}. We
note that since $\mathbb{E}\left[\partial_{t}S\right]=0$, the first
terms in Eqs. \eqref{eq: dS} and \eqref{eq: dF} do not contribute
to the mean rate.

\section{Cycle representation of kinematic NESS flux.}

\label{sec:cr}

The divergence-free stationary flux $\mathbf{J}^{*}$ can be expressed
in terms of a {\em bivector potential} $\mA(\vx)$ {for diffusion on $\mathbb{R}^n$}, $\nabla\times\mA=\mJ^{*}$.
Note that $\mA$ is also not a vector in $\mathbb{R}^{n}$ in general;
rather it is a bivector whose components $A_{ij}$ satisfy 
\begin{equation}
J_{i}^{*}(\vx)=\left(\nabla\times\mA\right)_{i}=\sum_{j=1}^{n}\partial_{j}A_{ij}(\vx).\label{vec-p}
\end{equation}
It is straightforward to verify that $\nabla\cdot\left(\nabla\times\mA\right)=0$.
See Appendix for the derivation. Throughout the paper, we fix the $\times$ notation in $\nabla\times\mA$
to denote the vector from a bivector potential $\mA$, $(\nabla\times\mA)_{i}\coloneqq\sum_{j=1}^{n}\partial_{j}A_{ij}$,
and the $\wedge$ notation in $\nabla\wedge\mathbf{u}$ to map a vector
$\mathbf{u}$ to a bivector: $(\nabla\wedge\mathbf{u})_{ij}\coloneqq\partial_{i}u_{j}-\partial_{j}u_{i}$.

\begin{figure}
\[
\includegraphics[scale=0.5]{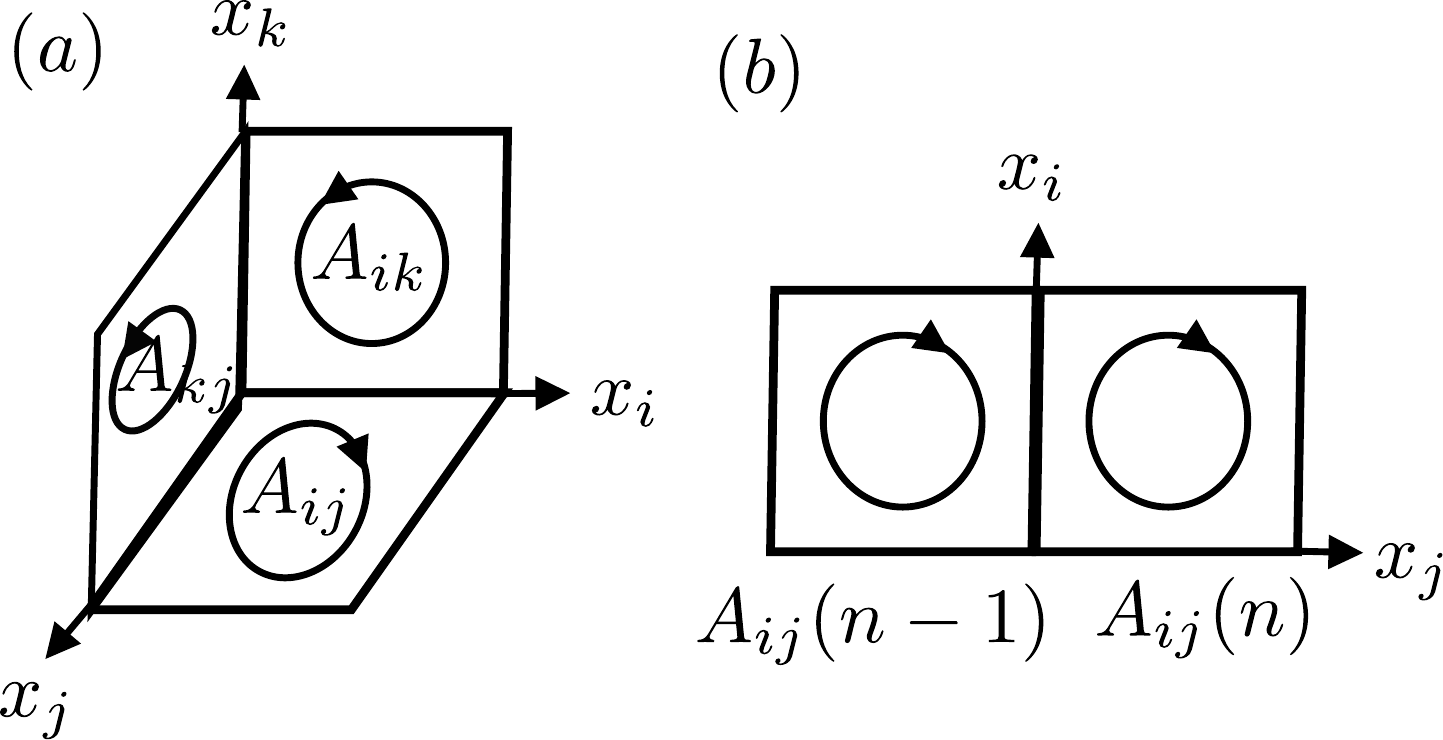}
\]
\caption{Bivector potential $A_{ij}$ and its derivative $\partial_{j}A_{ij}$
in a discrete lattice system. (a) For each unit vector $\hat{x}_{i}$
at a lattice point, there are $(n-1)$ number of $\hat{x}_{j}$ with
which $\hat{x}_{i}\wedge\hat{x}_{j}$ forms a bivector, an oriented
planar element. $A_{ij}$ is the cycle flux around the element. (b)
$A_{ij}(n)-A_{ij}(n-1)$ contributes to the edge flux along $\hat{x}_{i}$.
A similar $A_{ik}(m)-A_{ik}(m-1)$ also contributes to the same edge
flux. This is the geometric meaning of Eq. \eqref{vec-p}.}
\label{cycle flux} 
\end{figure}

{The physical meaning of Eq. \eqref{vec-p} is as followed. For every
infinitesimal change $\mathrm{d}x_{i}$ in the $x_{i}$ direction
at the point $\mathbf{x}\equiv(x_{1},\cdots,x_{n})\in\mathbb{R}^{n}$,
there are $(n-1)$ orthogonal directions $\mathrm{d}x_{j}$, $j\neq i$,
and $\mathrm{d}x_{i}\wedge\mathrm{d}x_{j}$ corresponds to a bivector,  an infinitesimal planar element in the $ij$-th plane, as shown in Fig. \ref{cycle flux}(a)}. Here, we give
the bivector potential $A_{ij}(\mathbf{x})$ a physical interpretation
as the \emph{stationary} \emph{cycle} \emph{flux} around the oriented
infinitesimal planar element $\mathrm{d}x_{i}\wedge\mathrm{d}x_{j}$
at $\mathbf{x}$. The $i$th component of $\mJ^{*}$, $J_{i}^{*}$,
is determined from all the neighboring infinitesimal planes $\mathrm{d}x_{i}\wedge\mathrm{d}x_{j}$,
$j\neq i$. $\partial_{j}A_{ij}$ contributes a net edge flux along
$x_{i}$ due to the pair of $A_{ij}$ at $(x_{1},\cdots,x_{j},\cdots,x_{n})$,
and at $(x_{1},\cdots,x_{j}-\mathrm{d}x_{j},\cdots,x_{n})$ as shown
in Fig. \ref{cycle flux}(b). An increasing $A_{ij}$ in the $j$th direction
leads to a positive net flow $\partial_{j}A_{ij}$ in the $x_{i}$
direction. In a word, Eq. \eqref{vec-p} gives a cycle representation
of the steady state fluxes $J_{i}^{*}$ along the edges in terms of
the cycle fluxes around the planar elements. An earlier discussion
for 3-D cases can be found in \cite{qian_vector_1998}. $\mA(\vx)$
is a potential of $\mJ^{*}(\vx)$ in terms of vorticity componenets
at $\vx$. 

\section{Landauer-Bennettt-Hill principle for diffusion.}

In NESS, the $\mJ(\mathbf{x},t)$ in Eq. \eqref{eq: average production rate}
is replaced by the divergence-free stationary flux $\mJ^{*}(\mathbf{x})$.
With our bivector potential, the mean rate of $\mathcal{W}$ in Eq.
\eqref{eq: change in a work-like quantity} with corresponding force
$\vf$ can be rewritten {by integration by part, \label{eq: Wrate}
\begin{subequations}
\begin{align}
\dot{\mathcal{W}}^{*}= & \int_{\mathbb{R}^{n}}\mJ^{*}\cdot\vf\rd V=\int_{\mathbb{R}^{n}}\left(\nabla\times\mathbf{A}\right)\cdot\vf\rd V\label{eq:}\\
= & \int_{\mathbb{R}^{n}}\mA\cdot\left(\nabla\wedge\vf\right)\rd V\label{eq: cycle flux * cycle affinity of arbitraryW}
\end{align}
\end{subequations}
where the scalar product in Eq. \eqref{eq:} is between two vectors whereas the scalar product in Eq. \eqref{eq: cycle flux * cycle affinity of arbitraryW} is between two bivectors, defined as the half Frobenius product of their antisymmetric matrix components .
Importantly, Eq. \eqref{eq: cycle flux * cycle affinity of arbitraryW} gives the
mean rate of $\mathcal{W}$ at NESS a new cyclic representation: it
is the average of vorticity $\nabla\wedge\vf$, weighted by the cycle
flux $\mA$. }
This immediately implies that thermodynamic quantities with a gradient
force $-\nabla U$ would have zero mean rate in NESS since $\nabla\wedge\left(-\nabla U\right)=0$.
That includes all the functions $\Phi$, $S$, and $F$, implying
that the mean rates of $\mathcal{S}_{\mathrm{tot}}$, $\mathcal{Q}$,
and $\mathcal{Q}_{\mathrm{hk}}$ are all identical at NESS, $\dot{\mathcal{S}}_{\mathrm{tot}}^{*}=\dot{\mathcal{Q}}^{*}=\dot{\mathcal{Q}}_{\mathrm{hk}}^{*}.$
This has been termed as a ``gauge freedom'' \cite{feng_potential_2011,polettini_nonequilibrium_2012}.

Thus, the average total entropy production rate at NESS can be written
as 
\begin{align}
\dot{\mathcal{S}}_{\mathrm{tot}}^{*}= & \int_{\mathbb{R}^{n}}\mA\cdot\Big[\nabla\wedge(\mD^{-1}\vb)\Big]\rd V.\label{eq: cycle flux * cycle affinity}
\end{align}
The stationary cycle flux $\mathbf{A}$ is a purely kinematic concept
that doesn't have any thermodynamic content. A closed loop $\Gamma$
in $\mathbb{R}^{n}$ contains a surface $\Sigma$ which can be tiled
by an array of tiny oriented infinitesimal planar elements at $\vx$,
for all $\vx\in\Sigma$. $\mA(\vx)$ then decomposes $\mJ^{*}(\vx)$,
following Kirchhoff's law, in terms of the occurrence rate of these
tiny oriented elements along the infinitely long, ergodic path $\vX_{t}$.
As a vorticity description of the NESS, $\mA$ is \textbf{\emph{nonlocally}}
determined.

On the other hand, as hinted by Eq. \eqref{eq: stochastic Stot in NESS}
and further by Eq. \eqref{eq: cycle flux * cycle affinity}, the \emph{cycle}
\emph{affinity} \cite{qian_entropy_2016}, as the Onsager's thermodynamic
force corresponding to the cycle flux, is \textbf{\emph{locally}}
determined through $\nabla\wedge\left(\mD^{-1}\vb\right)$. $\mD^{-1}\vb$
should be identified as the vector potential of the cycle affinity.
This is in sharp contrast to the standard expression $\dot{\mathcal{S}}_{\mathrm{tot}}^{*}=\int_{\mathbb{R}^{n}}\mJ^{*}\cdot\mD^{-1}\mathbf{v}^{*}\mathrm{d} V$
where the thermodynamic force corresponding to the edge flux $\mJ^{*}$is
nonlocally defined by $\mD^{-1}\vv^{*}$. Note that the cycle affinity
has components 
\begin{equation}
\left[\nabla\wedge\left(\mD^{-1}\vb\right)\right]_{ij}=\partial_{i}(\mD^{-1}\vb)_{j}-\partial_{j}(\mD^{-1}\vb)_{i},\label{eq: cycle affinity}
\end{equation}
representing how the \emph{two} dimensions $x_{i}$ and $x_{j}$ are
coupled.

This constitutes the LBH principle for diffusion processes: Entropy
production in NESS is characterized by the locally-defined cycle affinity
as a bivector; entropy production of a bigger loop is the integral
of the cycle affinity of infinitesimal cycles as shown in Eq. \eqref{eq: stochastic Stot in NESS};
and the average entropy production rate is the average cycle affinity,
weighted by the cycle flux of infinitesimal cycles as shown in Eq.
\eqref{eq: cycle flux * cycle affinity}. The fundamental unit of
NESS is the non-detailed-balanced kinetic cycle \cite{hill_free_2012},
in terms of bivectors.

\section{Mean rate decomposition outside of NESS.}

For the mean rate of thermodynamics quantities outside of NESS, we
rewrite Eq. (\ref{eq: average production rate}) as $\dot{\mathcal{W}}=\mathbb{E}\left[\mathbf{v}\left(\mathbf{X}_{t},t\right)\ \cdot\vf\left(\mathbf{X}_{t},t\right)\right]$.
By $\mathbf{v}(\vx,t)=\mathbf{v}^{*}(\vx)-\mD\nabla F(\vx,t),$ the
mean rate has a decomposition, 
\begin{equation}
\dot{\mathcal{W}}=\mathbb{E}\left[\vf\cdot\mathbf{v}^{*}\right]-\mathbb{E}\left[\vf\cdot\mD\nabla F\right].\label{eq: h decomposition}
\end{equation}
Outside of NESS, the two terms within can be rewritten as
{
\begin{subequations}\label{two alternate expression}
\begin{align}
\mathbb{E}\left[\vf\cdot\mathbf{v}^{*}\right] & =\int_{\mathbb{R}^{n}}\mA\cdot\nabla\wedge\left(\frac{p}{\pi}\vf\right)\rd V \label{eq: E f gamma*}\\
\mathbb{E}\left[\vf\cdot\mD\nabla F\right] & = - \int_{\mathbb{R}^{n}}\frac{p}{\pi}\nabla\cdot\left(\pi\mD\vf\right)\rd V.\label{eq:E D del F}
\end{align}
\end{subequations}
}
This implies an average perpendicularity between
$-\nabla F$ and $\mathbf{v}^{*}$, $\mathbb{E}[-\nabla F\cdot\mathbf{v}^{*}]=0.$ 

If with inner product defined as $\left\langle \mathbf{u},\mathbf{v}\right\rangle =\mathbb{E}\left[\mathbf{u}\cdot\mD^{-1}\mathbf{v}\right]$,
this average perpendicularity becomes $\left\langle -\mD\nabla F,\mathbf{v}^{*}\right\rangle =0$.
The decomposition in Eq. \eqref{eq: h decomposition} has a geometric
interpretation under the provided inner product. The mean rate of
a ``work''-like quantity $\mathcal{W}$ of Eq. \eqref{eq: change in a work-like quantity}
with the force $\vf$ is determined by the inner product of $\mD\vf$
with two perpendicular vectors, $\mathbf{v}^{*}$ and $-\mD\nabla F$,
\begin{equation}
\dot{\mathcal{W}}=\left\langle \mD\vf,\mathbf{v}^{*}\right\rangle +\left\langle \mD\vf,-\mD\nabla F\right\rangle .\label{eq: h inner product decomposition}
\end{equation}
This gives the Pythogarean-like relation between $\mathbf{v}$, $\mathbf{v}^{*}$,
and $-\mD\nabla F$ \cite{qian_kinematic_2020} hidden behind the
famous entropy production rate decomposition \cite{ge_extended_2009},
\begin{align}
\underbrace{\left\langle \mathbf{v},\mathbf{v}\right\rangle }_{\dot{\mathcal{S}}_{\mathrm{tot}}} & =\underbrace{\left\langle \mathbf{v}^{*},\mathbf{v}^{*}\right\rangle }_{\dot{\mathcal{Q}}_{\mathrm{hk}}}+\underbrace{\left\langle -\mD\nabla F,-\mD\nabla F\right\rangle }_{\dot{F}_{\mathrm{d}}}.\label{eq: EP decomposision}
\end{align}
Results above indicate that $\mathbf{v}^{*}$ and $-\mD\nabla F$
originate from two rather disjoint irreversibilities, and that the
geometry defined through the Riemannian metric $\mathbf{D}^{-1}(\mathbf{x})$
may be the most natural one in thermodynamics.

\section{Onsager's reciprocality and the Onsager operator.}

A diffusion process in $\mathbb{R}^{n}$ always has its NESS thermodynamic
force $\mathbf{D}^{-1}\mathbf{v}^{*}$ linearly related to transport
flux $\mJ^{*}$, $\left(\mathbf{D}^{-1}\mathbf{v}^{*}\right)_{i}$
$=e^{\Phi}\sum_{j=1}^{n}D_{ij}^{-1}J_{j}^{*}$. Many previous studies
have explored this unique feature \cite{qian_mesoscopic_2001,reguera_mesoscopic_2005}.
In the bivectorial representation there is a further linear affinity-vorticity
relationship 
\begin{align}
\nabla\wedge\left(\mD^{-1}\vb\right)(\vx) & =\mathcal{O}\mA(\vx).\label{eq: cycle affinity and flux}
\end{align}
where $\mathcal{O}=\nabla\wedge\left(e^{\Phi}\mD^{-1}\nabla\times\right)$.
We shall call the operator $\mathcal{O}$ the {\em Onsager operator}.
It linearly relates the cycle flux bivector to the cycle affinity bivector.
{As an example for the Onsager operator superimposing different $\mA$ from different processes,
we note that $(\Phi,\mathbf{A},\mathbf{D})$ together specify a diffusion process on $\mathbb{R}^n$ with $\Phi$ and $\mA$ as the scalar and bivector potentials of $\vb$.
Therefore, for a family of systems with fixed $\Phi$ and $\mD$, different bivector potential $\mA$ gives us different processes.
With $\mathcal{O}$ fixed in a family, a third process in this family with $\mA_3 = \mA_1 +\mA_2$ will have cycle affinity $\mathcal{O}\mA_3 = \mathcal{O}\left(\mA_1 +\mA_2\right)=\mathcal{O}\mA_1 +\mathcal{O}\mA_2$.   With in this family, $(\Phi,\mathbf{0},\mathbf{D})$ is a reversible process; and $(\Phi,-\mathbf{A},\mathbf{D})$ is 
the adjoint process.
}

\begin{figure}
\[
\includegraphics[scale=0.3]{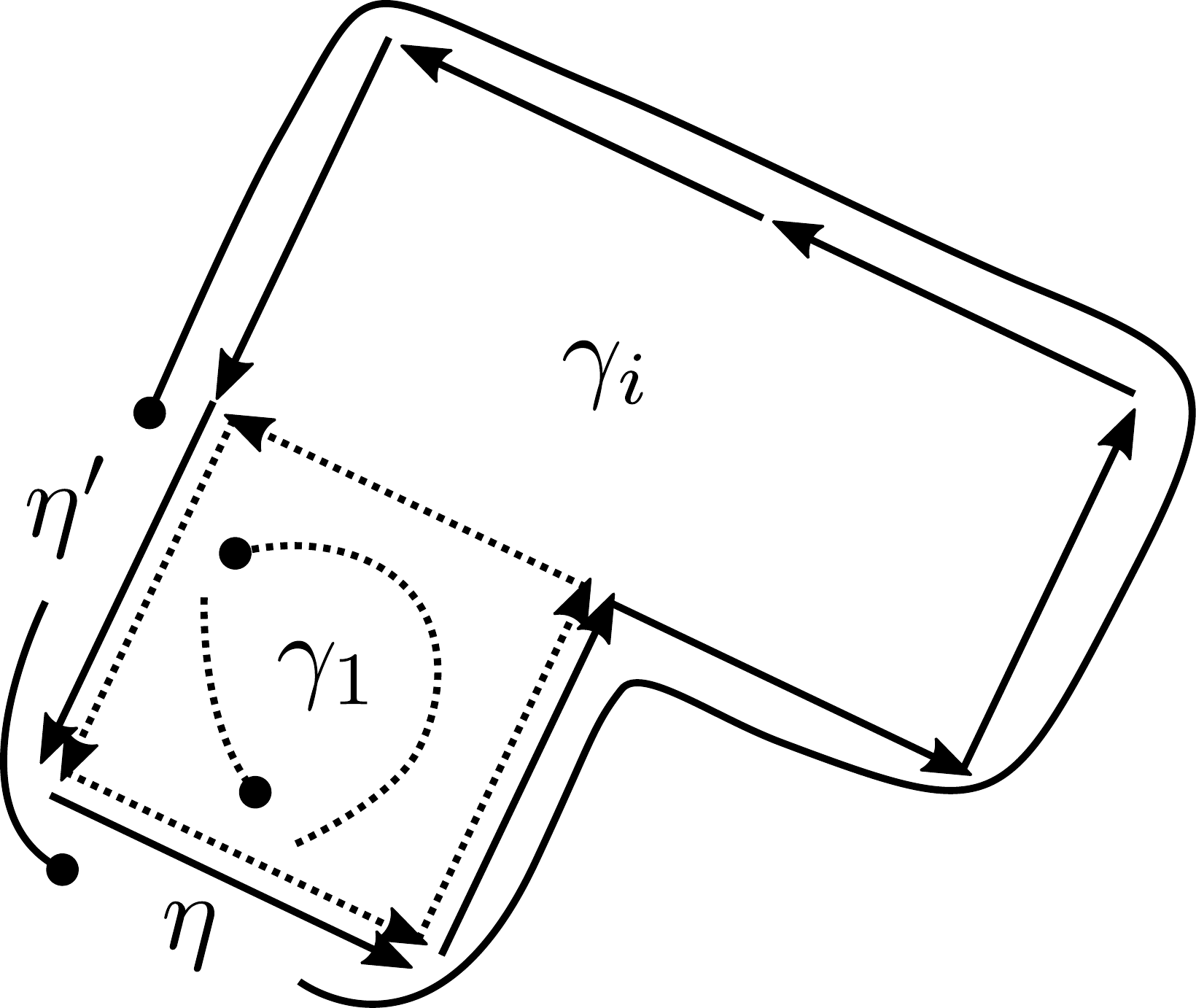}
\]
\caption{{Two edges $\eta$ and $\eta'$ are coupled through all the cycles $\gamma$ they are both in. The cycle-coupling is the origin of Onsager's reciprocality. Arrows represent cycle edges, and line with dot at the end represents the driving from force to flux. Dash lines indicate the smallest cycle coupling and solid lines indicate another cycle coupling. }}
\label{fig: edge coupled through cycle} 
\end{figure}

For discrete-state systems, it was understood in \cite{hill_linear_1982,qian_entropy_2016}
that such a relation at the cycle level is the fundamental origin
of the {Onsager's reciprocality and symmetry.
Specifically, if we denote $G_{\gamma \eta}$ as the incidence matrix of an edge $\eta $ in a cycle $\gamma$, then the stationary net edge flux is coupled with other edges through cycles,
\begin{equation}
J_{\eta} = \sum_{\gamma \in \mathrm{all\ cycles}} G_{\gamma \eta} J_{\gamma}^{-} \left[\exp\left(\sum_{\eta'}G_{\gamma \eta'}\mathcal{Q}(\eta')\right)-1 \right] \label{Onsager reciprocality in discrete}
\end{equation}
where $J_{\gamma}^{-}$ is the one-way cycle flux with orientation reversed to $\gamma$.
The reciprocality between $\eta$ and $\eta'$ manifests in that the mutual interaction between them are coupled only through cycles that contain both $\eta$ and $\eta'$:
Among all cycles, only those which contains $\eta$ contribute to $J_{\eta}$; among them, only those which contains $\eta'$ has the effect from $\eta'$, \emph{i.e.} a $\mathcal{Q}(\eta')$ term in $(\cdots)$; and only among these shared cycles does $\mathcal{Q}({\eta})$ affect back to $J_{\eta'}$, as illustrated by Fig. \ref{fig: edge coupled through cycle}.
The symmetry of the reciprocality further requires the system to be near equilibrium in general discrete-systems. With $J^{\mathrm{eq}}_{\gamma}$ denoting the equilibrium one-way flux, $J_{\gamma}^{-}\simeq J^{\mathrm{eq}}_{\gamma}$, and the cycle affinity to be small near equilibrium $\mathcal{Q}_\gamma\ll 1$ for all cycle $\gamma$, Eq. \eqref{Onsager reciprocality in discrete} becomes, to-the-leading-order,
\begin{equation}
J_{\eta} \simeq \sum_{\eta'} \left(\sum_{\gamma \in \mathrm{all\ cycles}} G_{\gamma \eta} J^{\mathrm{eq}}_{\gamma} G_{\gamma \eta'} \right) \mathcal{Q}_{\eta '} \label{Onsager in discrete edge}
\end{equation}
with symmetric Onsager matrix given by $\left(\cdots\right)$.

The bivectorial cycles in diffusion processes demonstrates vividly the reciprocality between two dimensions. How treating a diffusion process as a discrete-state Markov process on infinitesimal lattice system could reveal the fundamental origin of the general force-flux linearity and the symmetry in the Onsager's reciprocality for systems even far away from equilibrium remain to be ellucinated in future works.} 
We note that the mean NESS entropy production rate has a simple bilinear form: 
\begin{align}
\dot{\mathcal{S}}_{\mathrm{tot}}^{*} & =\int_{\mathbb{R}^{n}}\mA\cdot\mathcal{O}\mA\rd V.\label{eq: Onsager operator}
\end{align}
Incidentally, Onsager also considered tiny vortices ``who wanted
to play'' as the fundamental objects in hydrodynamic turbulent flow
\cite{eyink_onsager_2006}.

\section{The probabilistic gauge of the bivector potential $\mA$.}

The bivector potential $\mA$ obtained for the divergence-free $\mJ^{*}(\vx)=\nabla\times\mA(\vx)$
is not unique: It has a gauge freedom with an arbitrary curl-free
bivector. The situation has an analogue to that of discrete-state
Markov process \cite{qian_decomposition_1979,qian_circulation_1982},
and the vector potential in classical electrodynamics. Interestingly,
for discrete-state Markov process, Qian and Qian have proven the existence
and uniqueness of a gauge with a probabilistic meaning: Cycles are
not just represented in terms of Kirchhoff decomposition via linearly
independent bases; rather the space of {\em all possible cycles}
are considered, on which the unique \emph{probabilistic gauge}, as
NESS cycle flux, is the occurrence rate of a given cycle along the
infinitely long, ergodic path \cite{qian_circulation_1982}. Whether
such a unique probabilistic gauge also exists for bivector potential
$\mA$ on $\mathbb{R}^{n}$, or a more extended space of loops \cite{lejan_2010},
remains to be further investigated.

\section{Conclusions and discussion.}

This study clearly points to the importance of {\em cycle representation}
for mesoscopic nonequilibrium thermodynamics (NET) in terms of cycle
flux $\mA$ and cycle affinity $\nabla\wedge\big(\mD^{-1}\vb\big)$.
The former is a pure kinematic concept and the latter contains all
the fundamental information on NET. We show that the cycle flux and
cycle affinity are not simple vectors in $\mathbb{R}^{n}$; rather
they are bivectors, which can be represented by skew-symmetric matrices.
The cycle flux is the bivector potential of the conventional NESS
flux; and the cycle affinity has a vector potential $\mD^{-1}(\vx)\vb(\vx)$
which is obtained \emph{locally}.

Some of the mathematics in the present work is contained in the diffusion
process on a manifold \cite{qian_entropy_1999} and the gauge field
formulations of NET \cite{feng_potential_2011,polettini_nonequilibrium_2012}.
The present work provides a clearer physics of NET in phase space
as a formulation of Onsager's general principle for entropy production.
We identify the bivector nature of the cycle representation in terms
of a local cycle affinity and a nonlocal kinematic cycle flux; and
reveal a unified Landauer-Bennett-Hill thermodynamic principle for
stationary nonequilibrium systems.

Finally, we noted a parallel between quantum mechanical phase giving
a reality to the ``indeterminate'' vector potential in electromagnetism
\cite{wu_evolution_2006} and our stochastic formulation giving a
vorticity intepretation to the bivector $\mA$ in stochastic thermodynamics:
Steady state flux $\mJ^{*}$ turns out to be a derivative.

\begin{acknowledgements}
The authors thank Yu-Chen Cheng, Hao Ge, Hans C. \"{O}ttinger,
Matteo Polettini, David A. Sivak, and Jin Wang
for helpful feedback and discussions. The second author acknowledges Profs. Zhang-Ju Liu (PKU) and Xiang Tang (Wash.
U.) for teaching him the mathematics. {We also thank the two anonymous reviewers for their helpful suggestions}.
\end{acknowledgements}

%
\section*{Conflict of interest}
%
This work is partially supported
by the Olga Jung Wan Endowed Professorship for the second author. The authors have no conflicts of interest to declare that are relevant to the content of this article.

\section*{Appendix}
\setcounter{equation}{0}
\renewcommand{\theequation}{A \arabic{equation}}
{Here we summarize the mathematics used to derive the results in the present work. 
In the main text, we used the notion of multivariable calculus and the notion of wedge product without the introduction of differential form for simplicity. 
However, the concept of differential form and the associated exterior calculus are needed to derive the formula of generalized curl
and cross product for dimensions higher than $3$ \cite{arnold_mathematical_1997,fortney_visual_2018}. 
We shall introduce and use the differential form calculus here. Throughout the text, we will use Cartesian coordinate to describe the entire Euclidean $\mathbb{R}^n$.

\subsection*{A. Differential form and Integration} \label{Appendix A}
In vector calculus, the infinitesimal work done by a force $\vf$ from time $t$ to $t+\delta t$ on a path $\vy(t)$ is given by 
\begin{equation}
\delta \mathcal{W}=\vf(\vy(t)) \cdot \delta \vy(t)=\sum_{j=1}^n f_j(\vy(t)) \delta y_j(t+\delta t) \label{dW in vector}
\end{equation}
where $\delta \vy(t)$ denotes the infinitesimal vector $\vy(t+\delta t)-\vy(t)$. Usually, to emphasize the infinitesimal limit $\delta t \rightarrow 0$, we replace $\delta$ with $\rd$, leading to a notation $\vf(\vy(t)) \cdot \rd \vy(t)$.
However, in the mathematics of differential form, the operater $\rd$ is generalized and understood differently.
In the main text, we used $\rd$ as the standard infinitesimal difference operator in  calculus. Here, we shall use $\delta$ as the infinitesimal operator and $\rd$ as the \emph{exterior derivative} of differential form, as we will introduce below.

We first introduce the concept of \emph{1-form}s, which are linear functions that maps a vector to a real number. 
Notice that the infinitesimal work in Eq. \eqref{dW in vector} actually takes the (tangent) vector $\delta \vy(t)$ at a point $\vy(t)$ and return us a number. The infinitesimal work is thus generally a differential 1-form at a given point, say $\boldsymbol{\xi}$, associated with a force vector $\vf=(f_{1},f_{2},\cdots,f_{n})$,
\begin{equation}
\omega_{\boldsymbol{\xi}}(\vu)=\sum_{j=1}^{n}f_{j} (\boldsymbol{\xi}) \rd x_{j}(\vu).\label{1-form}
\end{equation}
It takes an infinitesimal vector $\vu$ as an input and gives us the infinitesimal work generated when going from $\boldsymbol{\xi}$ to $\boldsymbol{\xi}+\vu$.
The basis of a 1-form is $\{\rd x_{j}\}$, which are themselves 1-forms\footnote{Here, just treat $\rd x_{j}$ as an notation as the basis of a 1-form.}: 
$\rd x_i$ takes a vector $\vu$ and gives us its $i$th component,

\begin{equation}
\rd x_i \left(\sum_j u_j \mathbf{e}_j\right)=\mathbf{e}_i\cdot \left(\sum_j u_j \mathbf{e}_j\right)=u_i
\end{equation}
where $\mathbf{e}_i$ is the unit vector in the $i$th direction.
To match up Eq. \eqref{dW in vector} with Eq. \eqref{1-form}, simply take $\vu \coloneqq \delta \vy(t)$ and $\boldsymbol{\xi}=\vy(t)$. The relation $\rd x_i (\square) = \mathbf{e}_i\cdot \square$ is what allows us to write the differential form in a vectorized expression in the main text,
\begin{subequations}
\begin{align}
\omega_{\boldsymbol{\xi}}(\vu) &=\sum_i f_i(\boldsymbol{\xi})\rd x_i \left(\sum_j u_j \mathbf{e}_j\right)\\
&=\sum_i f_i(\boldsymbol{\xi})\mathbf{e}_i\cdot \left(\sum_j u_j \mathbf{e}_j\right)=\vf(\boldsymbol{\xi}) \cdot \vu.
\end{align}
\end{subequations}

A differential form is what we can integrate over a manifold. The integral of the work over a path $\Gamma =\{\vy(s), 0\le s \le t\}$ is then 
\begin{equation}
\int_{\Gamma} \omega= \int_{\Gamma} \sum_{j=1}^n f_j \rd x_j = \int_{0}^t \sum_{j=1}^{n}f_{j} (\boldsymbol{\vy}(s)) \rd x_{j}(\delta \vy(s)) \label{work integral parameterized}
\end{equation}
where inputs of the differential form are suppressed concisely before the parameterization in the last step. 
To carry out the computation, one would proceed with $\rd x_{j}(\delta \vy(s))=\delta \vy_j(s)=\vy_j'(s)\rd s$, which makes Eq. \eqref{work integral parameterized} an usual one dimensional integration w.r.t. $s$. 
This identification of differential form turns out to be significant for the general integral of $m$-form on a general manifold and the generalization of Stoke's theorem.

\subsection*{B. Bivector and 2-form} \label{Appendix B}
Stokes theorem in $\mathbb{R}^3$ tells us that a line integral of a vector field on a closed loop is equal to the surface integral of the vector field's curl. 
The intuition behind is that the curl of the vector field gives the vorticity of the vector field of an infinitesimal plane object. 
When integrating all the infinitesimal planes that tile the surface, neighboring circulation cancels and all the vorticity of the infinitesimal plane combines to give the vorticity on the big loop on the boundary.
This intuition is still valid in $\mathbb{R}^n$.
To see that, we shall first introduce how the ``planary object'' is represented in general $\mathbb{R}^n$: it is given by the notion of a (simple) bivector.

Two parameters are needed to parametrize a surface, and a surface can be cut into infinitesimal two dimensional parallelograms with edges given by the two infinitesimal tangent vectors of a point. 
Moreover, circulation of a vector field over an infinitesimal plane can have two orientations.
Putting these together, we use the notion $\vu \wedge \vv$ to represent an oriented parallelogram object spanned by two vectors $\vu$ and $\vv$, thus the name \emph{bivector}.
 The orientation of the object is reflected by the anti-symmetry of the wedge product $\wedge$, $\vu \wedge \vv=-\vv \wedge \vu$. 
The wedge product $\wedge$ is a linear operation satisfying $(c_1 \vu_1+c_2 \vu_2)\wedge \vv = c_1 \vu_1\wedge \vv+ c_2 \vu_2 \wedge \vv$ where $c_1,c_2 \in \mathbb{R}$. With these, we can get the component form of the bivector 
\begin{equation}
\vu \wedge \vv = \sum_{1\le i<j \le n} (u_i v_j - u_j v_i) \mathbf{e}_i \wedge \mathbf{e}_j
\end{equation}
with basis $\mathbf{e}_i \wedge \mathbf{e}_j$. Importantly, one can show that the area of the parallelogram, denoted as $\Vert \vu \wedge \vv \Vert$ is given by
\begin{equation}
\Vert \vu \wedge \vv \Vert^2 = \sum_{1\le i<j \le n} (u_i v_j - u_j v_i)^2.
\end{equation}
An inner product between two bivectors $\mathbf{A}=\sum_{i<j} A_{ij} \mathbf{e}_i \wedge \mathbf{e}_j$ and $\mathbf{B}=\sum_{i<j} B_{ij} \mathbf{e}_i \wedge \mathbf{e}_j$ is thus naturally defined as 
\begin{equation}
\mathbf{A}\cdot \mathbf{B} = \sum_{1\le i<j \le n} A_{ij} B_{ij} \label{inner product of two bivectors}
\end{equation}
with $\left(\mathbf{e}_i \wedge \mathbf{e}_j\right)\cdot\left(\mathbf{e}_k \wedge \mathbf{e}_l\right)=\delta_{ik} \delta_{jl}$ for $i<j$ and $k<l$. 

We note that, in general, bivectors are objects that can be expressed as $\mathbf{A}=\sum_{i<j} A_{ij} \mathbf{e}_i \wedge \mathbf{e}_j$. Not all bivector can be expressed as the wedge product of two vectors. 
Such bivectors are called \emph{simple bivector}s, and only simple bivectors can have the geometrical meaning as a  parallelogram spanned by two vectors:
a general bivector can be the sum of many simple bivectors, superposition of many parallelogram. 
We also note that due to the anti-symmetry of the wedge product, the component $A_{ij}$ of a bivector $\mA$ can be represented by an anti-symmetric matrix. Then, the inner product between two bivectors, as shown in Eq. \eqref{inner product of two bivectors}, is the half of the Frobenius product of their anti-symmetric components.

   Now, similar to a 1-form taking a vector to a real number, a 2-form takes a bivector to a real number.The basis of a 2-form is given by $\{\rd x_i \wedge \rd x_j\}$ for $ 1\le i <j \le n$. Specifically, for $1\le i<j \le n$,
\begin{equation}
\rd x_i \wedge \rd x_j \left(\sum_{k<l} A_{kl} \mathbf{e}_k \wedge \mathbf{e}_l \right) = \left(\mathbf{e}_i \wedge \mathbf{e}_j \right)\cdot \left(\sum_{k<l} A_{kl} \mathbf{e}_k \wedge \mathbf{e}_l \right) = A_{ij}.
\end{equation}
Again, the relation between $\rd x_i \wedge \rd x_j \left(\mathbf{A} \right) = \left(\mathbf{e}_i \wedge \mathbf{e}_j \right)\cdot \mathbf{A}$ is what allow us to rewrite a differential 2-form in a vectorized form in the main text,
\begin{subequations}
\begin{align}
\omega(\mathbf{A}) &=\sum_{i<j}B_{ij}\rd x_i \wedge \rd x_j \left(\mathbf{A} \right) = \sum_{i<j}B_{ij}\left(\mathbf{e}_i \wedge \mathbf{e}_j \right)\cdot \mathbf{A}\\
&= \sum_{i<j} B_{ij} A_{ij}=\mathbf{B}\cdot \mathbf{A}.
\end{align}
\end{subequations}
When integration the differential 2-form over a surface, the $\mathbf{A}$ as the input of the 2-form here would be the infinitesimal bivector given by the two infinitesimal tangent vectors at a point, representing the infinitesimal tangent parallelogram at the point.

\subsection*{C. Exterior Derivative and the Curl of a vector field}\label{appendix C} 
The concept of curl in vector calculus is useful because of the Stokes theorem in $\mathbb{R}^3$. We shall thus use the generalized version of the Stokes theorem, the Stoke-Cartan theorem, to motivate the notion of \emph{exterior derivative} and get the general definition of the curl of a vector field in $\mathbb{R}^n$.

The Stokes-Cartan theorem states that the integral of a differential
form $\omega$ over the boundary of some oriented manifold $\Omega$
is equal to the integral of its exterior derivative $\rd\omega$ over
the whole of $\Omega$: 
\begin{equation}
\oint_{\partial\Omega}\omega=\int_{\Omega}\rd\omega.\label{S-C}
\end{equation}
In a sense, the exterior derivative $\rd$ is defined so that Eq. \eqref{S-C} holds for a general manifold.
We shall thus understand it with Stokes-Cartan theorem: the exterior derivative of a differential form $\omega$ can be interpreted,
geometrically, as the integral over the boundary of an infinitesimal
parallelepiped $h\Omega$, 
\begin{equation}
\rd \omega = \lim_{h\rightarrow 0} \frac{1}{\Vert h \Omega\Vert} \int_{\partial (h \Omega)} \omega
\end{equation}
where $\Vert h \Omega\Vert$ denotes the volume of $h\Omega$.
With this, one sees that the twice exterior derivative of any differential form has to be zero, $\rd \rd \omega =0$. This is by applying the Stokes-Cartan theorem twice for a form that is itself the exterior derivative of another form $\varphi=\rd \omega$ (such form is called \emph{exact}). For an arbitrary compact region $\Omega$, we have 

\begin{equation}
\int_{\Omega}\rd \rd \omega=\int_{\Omega}\rd\varphi = \int_{\partial \Omega}\varphi = \int_{\partial \Omega}\rd \omega = \int_{\partial \partial \Omega} \omega.
\end{equation}
Since $\partial \partial \Omega=\emptyset$ and $\Omega$ is arbitrary, we have $\rd \rd \omega =0$. A form with zero exterior derivative is said to be \emph{closed}. Therefore, every exact form is closed.

The exterior derivative of a $k$-form is a $(k+1)$-form. The exterior derivative of a general form $\alpha \wedge \beta$ obeys the product rule: Suppose $\alpha$ is a $p$-form, then 
\begin{equation}
\rd \left(\alpha \wedge \beta \right)=\rd \alpha \wedge \beta +\left(-1\right)^p \left(\alpha \wedge \rd \beta\right).
\end{equation}
Applying this and $\rd \rd \omega =0$, one can get the exterior derivative of the 1-form in Eq. \eqref{1-form},
\begin{subequations}
\begin{align}
\rd \left(\sum_{j=1}^{n}f_{j}\rd x_{j}\right) &= \sum_{j=1}^n \left(\rd f_j\right) \wedge \rd x_j
= \sum_{1\le i,j\le n} \left(\frac{\partial f_{j}}{\partial x_i} \rd x_i\right) \wedge \rd x_{j} \\
&= \sum_{1\le i<j\le n} \left(\frac{\partial f_{j}}{\partial x_i} - \frac{\partial f_{i}}{\partial x_j} \right) \rd x_i \wedge \rd x_{j} 
\end{align}
\end{subequations}
where $\rd x_i \wedge \rd x_j=-\rd x_j \wedge \rd x_i$ is used. By Stokes-Cartan theorem, we the have
\begin{equation}
\oint_{\partial\Omega} \sum_{j=1}^{n}f_{j}\rd x_{j} =\int_{\Omega}\sum_{1\le i<j\le n} \left(\frac{\partial f_{i}}{\partial x_j} - \frac{\partial f_{i}}{\partial x_j} \right) \rd x_i \wedge \rd x_{j}.\label{S-C detailed}
\end{equation}
As introduced in Sec. \hyperref[Appendix B]{B}, we can rewrite Eq. \eqref{S-C detailed} in the vectorized form, 
\begin{equation}
\oint_{\partial\Omega} \vf \cdot \rd \vx =\int_{\Omega} \nabla \wedge \vf \cdot \rd \boldsymbol{\sigma}\label{S-C for bivector}
\end{equation}
where inner product between bivectors was introduced in Eq. \eqref{inner product of two bivectors}. Hence, $\nabla \wedge \vf$ as a bivector is the curl of the vector field $\vf$.

Since $\rd \rd \omega =0$, a gradient vector field is always curl-free, \emph{i.e.} $\nabla \wedge \nabla U=0$ where $U$ is a scalar potential. For the converse, we apply Poicar\'{e} lemma, which states that every closed form is exact (locally) on a contractible domain. Since we are concern with processes on the entire Euclidean manifold $\mathbb{R}^n$, which is contractible, we can conclude that curl-free vector field is globally a gradient field.

\subsection*{D. Bivector potential of a divergence-free vector field}\label{appendix C}
Using exterior derivatives and differential forms, the integral of
a vector field $\vF(\vx)$ over an $(n-1)$-dimensional
closed surface $\Sigma$ as flux is 
\begin{subequations}
\begin{align}
\oint_{\Sigma} \sum_{k=1}^{n}F_{k}(\vx)\rd\sigma_{k}
& =\int_{\Omega} \rd \left(\sum_{k=1}^{n}F_{k}(\vx)\rd\sigma_{k}\right)\\
& =\int_{\Omega}\sum_{k=1}^{n}\left[ \sum_{j=1}^{n}\left(\partial_{j}F_{k}\right)\rd x_{j}\right] \wedge \rd\sigma_{k} \\
 & =\int_{\Omega}\sum_{k,j=1}^{n}\left(\partial_{j}F_{k}\right)\rd x_{j}\wedge\rd\sigma_{k}\\
 & =\int_{\Omega}\sum_{j=1}^{n}\left(\partial_{j}F_{j}\right)\rd V,\label{eq:equationA21}
\end{align}
\end{subequations}
where $\Omega$ is the $n$-volume contained by the closed $(n-1)$-surface
$\Sigma$, and 
\begin{equation}
\rd\sigma_{k}=(-1)^{k-1} \rd x_{1}\wedge\rd x_{2}\wedge\cdots\wedge\rd x_{k-1}\wedge\rd x_{k+1}\wedge\cdots\wedge\rd x_{n}
\end{equation}
with $\rd x_{k}$ missing. The $(-1)^{k-1}$ factor is to ensure $\rd x_{j}\wedge\rd\sigma_{k}=\delta_{jk}\rd V$ where 
\begin{equation}
\rd V=\rd x_1 \wedge \rd x_2 \wedge \cdots \wedge \rd x_n 
\end{equation}
 is the infinitesimal $n$ dimensional volume element. The $(n-1)$-form $\rd \sigma_i$ is the \emph{Hodge dual} of the 1-form $\rd x_i$, often denoted as  $\rd \sigma_i = \star \rd x_i$.

Now if a vector field $\vF(\vx)$ is divergence free, \emph{i.e.} $\sum_{j=1}^{n}\left(\partial_{j}F_{j}\right)=0$, then  the $(n-1)$-form has a zero exterior derivative,
\begin{equation}
\rd \left(\sum_{k=1}^{n}F_{k}(\vx)\rd\sigma_{k}\right)=0.
\end{equation}
Poicar\'{e} lemma then guarantees that on a contractible domain,
\begin{equation}
\sum_{k=1}^{n}F_{k}(\vx)\rd\sigma_{k}=\rd\omega,\label{n-1 form is exact}
\end{equation}
where $\omega$ is expected to be a $(n-2)$-form with the general
expression 
\begin{equation}
\omega=\sum_{1 \le i<j \le n}u_{ij}(\vx)\rd\eta_{ij},
\end{equation}
in which
\begin{equation}
\rd\eta_{ij}=(-1)^{i-1+j-2}\rd x_{1}\cdots\rd x_{i-1}\wedge\rd x_{i+1}\wedge\cdots\rd x_{j-1}\wedge\rd x_{j+1}\cdots\rd x_{n}
\end{equation}
with $\rd x_{i}$ and $\rd x_{j}$ missing. The $(-1)^{i-1+j-2}$ factor is to ensure $\left(\rd x_i \wedge \rd x_j\right)\wedge \rd \eta_{ij}=\rd V$ so that $\rd \eta_{ij} = \star \left(\rd x_i \wedge \rd x_j \right)$. 
Then, 
\begin{subequations}
\begin{align}
\rd\omega  & =  \sum_{1 \le i<j \le n}\sum_{k=1}^{n}\left(\partial_{k}u_{ij}\right)\rd x_{k}\wedge\rd\eta_{ij}\\
 & =\ \sum_{1 \le i<j \le n}\left\{ \left(\partial_{i}u_{ij}\right)\rd x_{i}\wedge\rd\eta_{ij}+\left(\partial_{j}u_{ij}\right)\rd x_{j}\wedge\rd\eta_{ij}\right\} \\
 & =\ \sum_{1 \le i<j \le n} (-1)\left(\partial_{i}u_{ij}\right)\rd\sigma_{j}+\sum_{1 \le i<j \le n}\left(\partial_{j}u_{ij}\right) \rd\sigma_{i} \\
 & =\ \sum_{1 \le j<i \le n} (-1)\left(\partial_{j}u_{ji}\right)\rd\sigma_{i}+\sum_{1 \le i<j \le n}\left(\partial_{j}u_{ij}\right) \rd\sigma_{i} \\
 & =\ \sum_{i=1}^{n}\sum_{j=1}^{n}\left(\partial_{j}A_{ij}\right)\rd\sigma_{i}.\label{eq:equation dw}
\end{align}
\end{subequations}
In the last step we have introduced $A_{ij}(\vx)=u_{ij}(\vx)$
for $i<j$, $A_{ij}(\vx)= -A_{ji}(\vx)$ for $i>j$ and $A_{ii}(\vx)=0$.
It is easy to verify that 
\begin{equation}
F_{i}(\vx)=\sum_{j=1}^{n}\left(\partial_{j}A_{ij}(\vx)\right) \label{bivector potential in components}
\end{equation}
is a divergence free vector field: 
\begin{equation}
\nabla\cdot\vF(\vx)=\sum_{i=1}^{n}\partial_{i}F_{i}(\vx)=\sum_{i,j=1}^{n}\left(\partial_{i}\partial_{j}A_{ij}(\vx)\right)=0.
\end{equation}
The vector potential $\mA(\vx)$ of a divergence-free field is a bivector with
anti-symmetric matrix components. 
Since we consider diffusion on the whole $\mathbb{R}^n$, which is contractible. Eq. \eqref{n-1 form is exact} and so Eq. \eqref{bivector potential in components} are thus globally valid.  

In $\mathbb{R}^3$, a divergence free vector field has a vector potential through the same curl differential operator as the one we used to compute the vorticity of a vector field. 
For general $\mathbb{R}^n$, this is no longer true. To distinguish the two in general, we have used $\nabla \wedge$ as the curl operator that maps a vector field to its vorticity bivector, with $\wedge$ reminding us the result is a bivector. Here, we use $\nabla \times$ to denote the differential operator that links a divergence-free vector field to its bivector potential. Eq. \eqref{bivector potential in components} is then expressed as 
\begin{equation}
\mathbf{F}(\vx)=\nabla \times \mathbf{A}(\vx). \label{bivector potential in vector}
\end{equation}
The close relation between them can be seen by integration by part:
For a divergence-free vector field $\mathbf{F}=\nabla \times \mathbf{A}$, we have
\begin{subequations}
\begin{eqnarray}
 &  & \int_{\mathbb{R}^{n}}\left(\nabla \times \mathbf{A}\right)\cdot\vu\ \rd V \ =\int_{\mathbb{R}^{n}}\sum_{i,j}\partial_{j}A_{ij}u_{i}\rd V \\
 & = & \int_{\mathbb{R}^{n}}\sum_{i<j}A_{ij}\left(\partial_{i}u_{j}-\partial_{j}u_{i}\right)\rd V=\int_{\mathbb{R}^{n}} \mathbf{A}\cdot \left(\nabla \wedge \mathbf{u}\right)\rd V.\hspace{1.2cm}
 \label{curl}
\end{eqnarray}
\end{subequations}
}

\bibliographystyle{spmpsci}      
\bibliography{BivectorJSP.bib}   

\end{document}